# Giant Phonon Anomalies in the Proximate Kitaev Quantum Spin Liquid α-RuCl₃

Haoxiang Li[1], T. T. Zhang[2,3], A. Said[4], G. Fabbris[4], D. G. Mazzone[5,6], J. Q. Yan[1], D. Mandrus[1,7], G. B. Halász[1], S. Okamoto[1], S. Murakami[2,3], M. P. M. Dean[5], H. N. Lee[1] and H. Miao[1,†]

[1]Material Science and Technology Division, Oak Ridge National Laboratory, Oak Ridge, Tennessee 37831, USA

[2]Department of Physics, Tokyo Institute of Technology, Okayama, Meguro-ku, Tokyo 152-8551, Japan

[3]Tokodai Institute for Element Strategy, Tokyo Institute of Technology, Nagatsuta, Midori-ku, Yokohama, Kanagawa 226-8503, Japan

[4]Advanced Photon Source, Argonne National Laboratory, Argonne, Illinois 60439, USA

[5]Condensed Matter Physics and Materials Science Department, Brookhaven National Laboratory, Upton, New York 11973, USA

[6]Laboratory for Neutron Scattering and Imaging, Paul Scherrer Institut, CH-5232 Villigen, Switzerland

[7]Department of Materials Science and Engineering, the University of Tennessee at Knoxville, Knoxville, Tennessee 37996, USA

†Correspondence should be addressed to miaoh@ornl.gov.

## Abstract

The Kitaev quantum spin liquid epitomizes an entangled topological state, for which two flavors of fractionalized low-energy excitations are predicted: the itinerant Majorana fermion and the $Z_2$ gauge flux. It was proposed recently that fingerprints of fractional excitations are encoded in the phonon spectra of Kitaev quantum spin liquids through a novel fractional-excitation-phonon coupling. Here, we detect anomalous phonon effects in α-RuCl₃ using inelastic X-ray scattering with meV resolution. At high temperature, we discover interlaced optical phonons intercepting a transverse acoustic phonon between 3 and 7 meV. Upon decreasing temperature, the optical phonons display a large intensity **enhancement near the Kitaev energy, $J_K \sim 8$ meV, that coincides with a giant acoustic phonon softening near the $Z_2$ gauge flux energy scale. These phonon anomalies signify the coupling of phonon and Kitaev magnetic excitations in $\alpha$-RuCl$_3$ and demonstrates a proof-of-principle method to detect anomalous excitations in topological quantum materials.**

**Introduction**

In correlated quantum materials, the nature of electronic interactions and their ground state topology is intimately linked to the geometry of the underlying lattice[1-6]. The low-energy excitations arising from pure electronic degrees of freedom inevitably interact with the crystal lattice, leaving behind their fingerprints in the phonon spectrum. Hitherto, the interactions of phonons with "conventional" quasiparticles of either Bose-Einstein or Fermi-Dirac statistics, such as magnons in magnets[7], phasons and amplitudons in density waves[8-10] and Bogliubons in superconductors[11], have been explored extensively. In contrast, the coupling between phonons and fractional excitations, including spinons in one-dimensional magnets[1-3,12-15], and Majorana fermions (MFs) and $Z_2$ gauge flux that are thought to exist in the Kitaev quantum spin liquids (QSL)[16-25], have remained elusive. The discovery of such fractional-excitation-phonon coupling (FPC) is of fundamental importance, as they carry key information of the intertwined quantum state[12-15]. In particular, the coupling of phonons to the itinerant MFs has been predicted to play a pivotal role in the realization of the field-induced quantum thermal Hall effect (QTHE) in $\alpha$-RuCl$_3$[26-29], which is a signature of quantum entanglement in Kitaev-QSLs[30-32].

Numerous studies have shown that the low-temperature phase of $\alpha$-RuCl$_3$ is a promising Kitaev-QSL candidate[17-24,30-32]. As displayed in Fig. 1a, the edge-sharing Ru-Cl octahedra form an effective spin-1/2 honeycomb network. The destructive quantum-interference through the close-to-90° Ru-Cl-Ru bonds significantly suppresses the Heisenberg magnetic exchange interaction, yielding a dominant Ising-type interaction (***J***) perpendicular to the Ru-Cl-Ru plane[33] (Fig. 1b). Figure 1c schematically depicts the phase diagram of α-RuCl$_3$. At zero magnetic field, the low-energy excitations in the paramagnetic phase are primarily determined by the Kitaev term[3,4,17-24,30-32]

$$H = \sum_{\gamma, <i,j>} J_K^\gamma S_i^\gamma S_j^\gamma \qquad (1)$$

Here $J_K^\gamma$ ($\gamma = X, Y, Z$) is the bond-dependent coupling parameter, and $<i,j>$ stands for nearest-neighbor pairs of spins at one of the *X, Y,* or *Z* bonds. The two characteristic energy scales are shown in Fig. 1d for the isotropic limit ($J_K^\gamma = J_K$). Below the Kitaev temperature scale $T_k \sim J_K$, the low-energy excitations of Eq. (1) start to fractionalize into itinerant MFs and fluctuating $Z_2$ gauge fluxes[34]. The former features a continuum that peaks broadly near $J_K$, while the latter is a local excitation with an energy around $0.065J_K$[16,20,25,34]. Below $T_N$=7 K, non-Kitaev interactions such as remnant Heisenberg magnetic exchange couplings, stabilize zigzag antiferromagnetic order that is suppressed under magnetic field[35-39]. Above **B**~7 T, a quantized thermal Hall conductivity (red region in Fig.1c) is observed, indicating strongly an entangled topological phase[30-32]. However, unlike the quantum Hall effect of electrons, it has been theoretically predicted that the QTHE can only be approximate and requires strong FPCs[26-29]. Here we report experimental signature of the FPC in α-RuCl$_3$ by uncovering two-types of phonon anomalies at zero magnetic field: a 35% enhancement of the phonon spectral weight near the Kitaev energy $J_K$, and a giant phonon softening of ~15% below 2 meV [26-29].

**Results**

Figure 2a shows the imaginary part of the dynamical phonon susceptibility $\chi''(\mathbf{Q}, \omega)$ along $\Gamma_1(6,-3,0)$-M(6,-2.5,0)-$\Gamma_2(6,-2,0)$ in reciprocal lattice units (r.l.u.) at room temperature (see Supplementary Note 6 for first-principles calculations of phonon dispersion). The dynamical susceptibility is given by the fluctuation-dissipation theorem via $\chi''(\mathbf{Q}, \omega) = S(\mathbf{Q}, \omega) \times (1 - e^{-\omega/k_B T})$, where $S(\mathbf{Q}, \omega)$ is the dynamical phonon structure factor that is directly measured by IXS. The total momentum transfer $\mathbf{Q}=\mathbf{q}+\mathbf{G}$, is composed of the reduced momentum transfer in the first Brillouin zone $\mathbf{q}$ and the reciprocal lattice vector $\mathbf{G}$. The elastic contribution at $\omega$=0 was subtracted by fitting the IXS raw data in the entire energy window (see Supplementary Note 1 and Note 2). We selectively probe in-plane transverse phonon modes, whose dispersions (open circles and open squares) and sinusoidal fits (dashed curves) are shown in Fig. 2b. As shown in Fig. 2a, the intensity of the transverse acoustic phonon changes significantly from $\Gamma_1$ to $\Gamma_2$, reflecting their different Bragg peak intensities that are plotted in Fig. 2c. Two low-energy optical phonons, $P_1$ and $P_2$, are observed at the Brillouin zone center $\Gamma_2$, corresponding to $\omega_1 = 2.7$ and $\omega_2 = 7$ meV, which are in good agreement with previous optical and neutron studies[19,40]. The two optical phonons carry opposite phonon velocities and form an interlaced structure that intercepts the acoustic phonon. An apparent phonon crossing occurs between $\Gamma$ and M (Fig. 2d and e), suggesting possible Dirac-cone and topological phononic nodal-lines[41-42].

In $\alpha$-RuCl$_3$, $J_K$ is estimated to be 5~9 meV in the low-temperature phase below 150 K[4,17-25,30] (more discussions in Supplementary Note 5), which roughly corresponds to the top of the $P_1$-$P_2$ phonon band. Thus, if Majorana-phonon coupling is present, phonon anomalies are expected in the energy range shown in Fig. 2b. Moreover, recent theoretical study of the pure Kitaev model predicts that the Majorana-phonon coupling is momentum dependent and peaked near the M and K point[28]. To uncover the energy and momentum dependent coupling between the optical phonons and the suggested MFs, we compare the temperature-dependent $\chi''(\mathbf{q}, \omega)$ along the M- $\Gamma_2$ path. A large spectral enhancement can be observed clearly in Fig. 3a-3f. Near the M point, the peak intensity of $P_1$ increases dramatically upon cooling from 300 K to 10 K. In contrast, the peak intensity of $P_2$ is unchanged except the 10 K data at the M point. When approaching the $\Gamma_2$ point (towards larger $|\mathbf{q}|$), the intensity enhancement is first decreases near the crossing-point ($P_1$ and $P_2$ crossed at $q=0.75$), but then reappears at $P_2$, which is higher in energy near the $\Gamma_2$ point. Interestingly, we find that the spectral enhancement is different between the symmetry related points $q=0.45$ and $q=0.55$. As we show in Fig. 2a, the transverse acoustic phonon starts to merge with the optical phonon near the M point. Since the acoustic phonon intensity is stronger at $q=0.45$, the asymmetric intensity enhancement suggests that the Majorana-phonon coupling is larger on the acoustic mode than the optical mode near $\omega \sim J_K$. To quantitatively show the spectral enhancement effect, we extract the temperature-induced difference in the integrated phonon intensity, $\Delta\chi''(\mathbf{q}, \omega_0) = \int_{\omega_0-\infty}^{\omega_0+\infty}[\chi''(\mathbf{q}, \omega, 10K) - \chi''(\mathbf{q}, \omega, 300K)]d\omega$, and plot $\Delta\chi''(\mathbf{q}, \omega_0)$ as function of $\Delta E = \omega_0 - \omega_{\max}$ in Fig. 3g. Here $\omega_0$ denotes the phonon peak position and $\omega_{\max} = 7$ meV is the band-top energy of $P_1$ and $P_2$. Unlike the broad continuum observed in the spin correlation function[17-19,21-23], $\Delta\chi''(\mathbf{q}, \omega_0)$ decreases rapidly as $\omega_0$ moves away from $J_K$ (Fig. 3). It also shows a strong momentum dependence with the enhancement occur around the high symmetry point M and $\Gamma_2$ (see Supplementary Fig. 9). This observation is in qualitative agreement with theoretical calculation that shows energy and momentum dependent Majorana-phonon coupling[28] (spectrum near the K point with spectral peak at higher energy is shown in Supplementary Fig. 6). The observed phonon enhancement is also consistent with a recent study of frustrated magnetic systems, which predicts large IXS cross-section for magnetic excitations[7].

We note, however, a quantitative understanding of the energy and momentum dependent optical phonon enhancement may require theoretical calculations beyond the pure Kitaev model.

We then turn to the transverse acoustic phonon near $\Gamma_1$. Figure 4a and b show the temperature-dependence of $\chi''(\mathbf{q}, \omega)$ at $\mathbf{q_1}=(0,0.1,0)$ (or $\mathbf{Q_1}=(6, -2.9, 0)$) and $\mathbf{q_2}=(0,0.15,0)$ (or $\mathbf{Q_2}=(6, -2.85, 0)$), respectively. At $\mathbf{q_1}$, the phonon peak position gradually shifts to lower energies. In contrast, it remains nearly unchanged at $\mathbf{q_2}$. The softening-effect is confirmed by directly comparing the raw data, $S(\mathbf{q}, \omega)$, at 10 and 300 K (Fig. 4c and d). The peak position is softened by about 13% at $\mathbf{q_1}$, which corresponds to ~0.3 meV shift in energy. Figure 4e and 4f show the relative peak shift $\omega_0(T)/\omega_0(300\ \mathrm{K})$ at $\mathbf{q_1}$ and $\mathbf{q_2}$ as function of temperature. We find that the acoustic phonon softening at $\mathbf{q_1}$ becomes progressively stronger below 80 K, consistent with the thermal Hall effect in $\alpha$-RuCl$_3$ where the thermal Hall conductivity, $\kappa_{xy}$, starts to increase. In Fig. 4e, we further show the phonon softening at $\mathbf{q_3}=(0,0.05,0)$. The error-bars returned from fittings are larger at $\mathbf{q_3}$ as the elastic intensity becomes stronger when approaching the Bragg peak. Interestingly, the relative phonon softening at $\mathbf{q_3}$ (~15%) is even larger when compared to $\mathbf{q_1}$. This suggests an enhanced renormalization for long wavelength acoustic phonons.

**Discussions**

The discovery of temperature and energy dependent phonon softening provides important information on the FPC in $\alpha$-RuCl$_3$. In the pure Kitaev model [Eq. (1)], quantum fractionalization occurs at $T_K \sim J_K \sim 100$ K [43], in agreement with our observations. Below $T_K$, the dispersionless gauge flux excitation crosses the linear dispersing acoustic phonon near $\omega = 0.065 J_K \sim 0.5$ meV [23] and induces a phonon anomaly near this energy scale (Fig. 4g). The observation of enhanced phonon softening [$\omega(\mathbf{q_1}) = 2$ meV and $\omega(\mathbf{q_3}) = 1$ meV] as $\omega \to 0.065 J_K$ is consistent with this picture, where the softening effect is expected to be significantly suppressed for $\omega(\mathbf{q}) \gg 0.065 J_K$. Figure 4h depicts another scenario that attempts to explain the phonon-softening. Here, the acoustic phonon and the itinerant MFs possess nearly identical linear dispersions at $\mathbf{q} \to 0$ [29]. This enhances Majorana-phonon coupling that yields a renormalization of the phonon dispersion below $T_K$ [28,29]. To justify this conjecture, we extract the acoustic phonon velocity $v_{ph} \sim 16$ meV·Å ($\hbar = 1$), which is based on the room-temperature phonon dispersion shown in Fig. 2. In the isotropic limit [16], the velocity of the itinerant MF is $v_{MF} = \frac{\sqrt{3}}{4} J_K a$, where the in-plane lattice constant $a = 5.9639$Å. Comparing $v_{ph}$ and $v_{MF}$ gives $J_K \sim 6.2$ meV, comparable to the experimental value. Besides the $Z_2$ gauge flux and MFs, in more realistic models with non-Kitaev interactions [44,45], other fractional excitations, such as spinons, may also be consistent with the observed phonon anomalies. It is important to note that the charge and magnetic excitations below 2 meV still remain unresolved in $\alpha$-RuCl$_3$. In particular, direct experimental evidence of $Z_2$ gauge flux is not well established yet. The observed acoustic phonon softening below 2 meV demonstrate a small energy scale in $\alpha$-RuCl$_3$ that strongly renormalize the acoustic phonon spectrum and hence may be responsible for the quantized thermal Hall effect.

Finally, we discuss the possibility of a magnon-phonon coupling. Below $T_N$, a gapped magnon excitation between 2~7 meV was observed in $\alpha$-RuCl$_3$ by previous neutron studies [19,21,22,37]. However, as we show in Figs. 3 and 4, the phonon anomalies onset at $T_K$, which is well above $T_N$. More importantly, evidence of an enhanced phonon softening is observed at $\omega = 1$ meV (see $\mathbf{q_3}$ in Fig. 4e and Supplementary Fig. 5), which is well below the magnon gap. Therefore, a magnon-phonon coupling is unlikely giving rise to the observed acoustic phonon softening. However, the magnon-phonon coupling may indeed present in α-RuCl₃. As we show in Fig. 2, the $P_2$ phonon energy is the same as the magnon energy near the M point[19,21,38]. Interestingly, the $P_2$ phonon intensity at the M point is enhanced at 10 K~$T_N$, supporting magnon-phonon coupling[46]. In addition, strong anharmonicity is proposed in the magnon excitation of this material[47], which represent the break-down of the spin quasiparticles. Such excitations contain extremely broad features[47] that is contradictory to the well-define energy scale of the phonon anomalies observed here.

Our discovery of two-types of phonon anomaly, *i.e.* the spectral enhancement in the optical phonon and the acoustic mode softening, provides experimental signature of FPC in the proximity of Kitaev-QSL[26,27]. Beyond the aforementioned implications, our observation has an even deeper impact on correlated topological quantum states. First of all, our approach can be immediately applied to other Kitaev-QSL candidates[1,3,4], such as iridates[4,48], where the inelastic neutron scattering experiments are difficult to perform due to strong neutron absorption of Ir. Moreover, it has been predicted that in $U(1)$ spin liquids the spinon Fermi surface features a large singularity at $2\mathbf{k}_F$, which induces phonon anomalies at $\mathbf{q}=2\mathbf{k}_F$ [12]. Both Kagome and triangular lattice have been speculated to host such charge neutral Fermi surfaces[49,50]. More recently, a giant thermal Hall effect has been observed in the cuprate high-$T_c$ superconductors[13] with large phonon contributions[51]. While mechanisms based on chiral spin liquid or topological spinons[14,15] have also been proposed, the theoretically predicted $\kappa_{xy}$ is 50% smaller than the experimental value[14], suggesting large phonon effect. Our observation of FPC in α-RuCl₃ validates phonons as a sensitive probe to uncover hidden fractional and non-local excitations, and hence can help to resolve key puzzles in correlated and entangled quantum states.

**Methods:**

**Sample preparation and characterizations:** Millimeter-sized $\alpha$-RuCl$_3$ crystals were grown by the sublimation of RuCl$_3$ powder sealed in a quartz tube under vacuum[52]. The growth was performed in a box furnace. After dwelling at 1060 ºC for 6 hours, the furnace was cooled to 800 ºC at 4 ºC/hour. Magnetic order was confirmed to occur at 7 K by measuring magnetic properties and specific heat[21].

**Inelastic x-ray scattering:** The experiments were conducted at beam line 30-ID-C (HERIX) at the Advanced Photon Source (APS). The highly monochromatic x-ray beam of incident energy $E_i$ = 23.7 keV (l = 0.5226 Å) was focused on the sample with a beam cross section of $\sim$35 × 15 mm$^2$ (horizontal × vertical). The total energy resolution of the monochromatic x-ray beam and analyzer crystals was $\Delta E \sim 1.3$ meV (full width at half maximum). The measurements were performed in transmission geometry. Typical counting times were in the range of 30 to 120 seconds per point in the energy scans at constant momentum transfer **Q**. H, K, L are defined in the trigonal structure with a=b=5.9639 Å, c=17.17 Å at the room temperature.

**Density functional theory calculation of phonon spectrum:** Phonon dispersions for $\alpha$-RuCl$_3$ were calculated using with density functional perturbation theory (DFPT) and the Vienna Ab initio Simulation Package (VASP). The exchange-correlation potential was treated within the generalized gradient approximation (GGA) of the Perdew-Burke-Ernzerhof variety, where the kinetic energy cutoff was set to 400 eV. Integration for the Brillouin zone were done by using a Monkhorst-Pack $k$-point grids which is equivalent to 8×8×9.

**Data availability:** The data that support the findings of this study are available from the corresponding author on reasonable request.

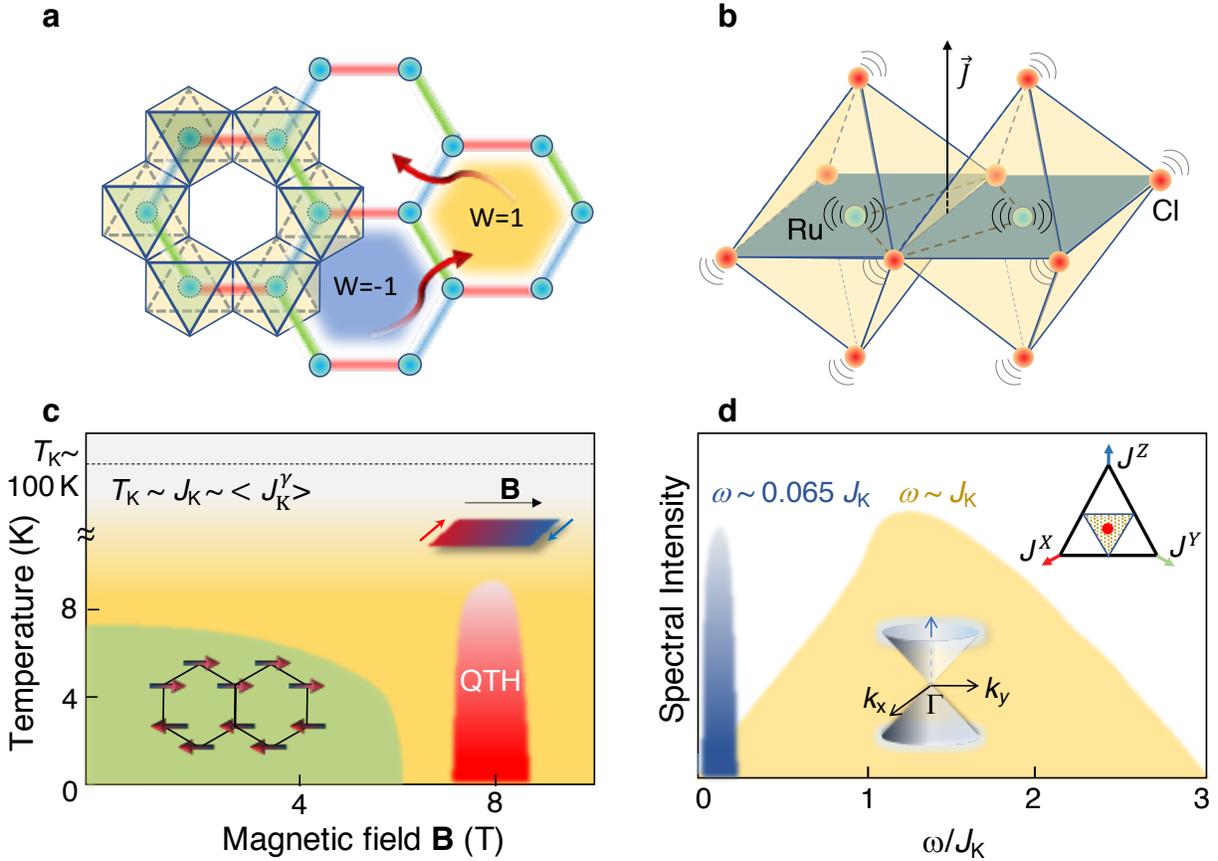

**Figure 1 Schematics of a Kitaev-QSL and the phase diagram of α-RuCl₃. (a)** Structure motif of α-RuCl₃ based on a honeycomb lattice of edge-sharing Ru-Cl octahedra. The red, green and blue bonds represent three orthogonal Kitaev interactions $J_K^\gamma$. In the pure Kitaev model [Eq. (1)], the low-energy excitations fractionalize into itinerant MFs (red arrows) and anyonic $Z_2$ flux (blue and yellow hexagons). **(b)** The nearly 90º Ru-Cl-Ru bonds and the moderate spin-orbit-coupling favor an Ising-type magnetic interaction that is perpendicular to the Ru-Cl-Ru plane highlighted in a blue-green-color. Lattice vibrations perturbatively modify the magnetic interactions, which induce a coupling between phonons and fractional excitations. **(c)** illustrates the phase diagram of α-RuCl₃ on a logarithmic-scale: below $T_K \sim J_K \sim 8$ meV (the yellow area) the thermal Hall conductivity, $\kappa_{xy}$, becomes finite, indicating a proximate Kitaev-QSL with MF and $Z_2$ gauge flux.

In the green area ($T<T_N$=7 K), non-Kaitaev terms drive the system into zigzag antiferromagnetic order. Under an external magnetic field (**B**>7 T) that completely suppresses the magnetic order, the system is driven into a QTHE state at finite temperature (the red area). **(d)** schematically shows two characteristic Kitaev energy scales in the isotropic limit: the itinerant MF excitations[18] (yellow area) that is broadly peaked around $J_K$ and the $Z_2$ gauge flux excitations (blue area) near 0.065 $J_K$.

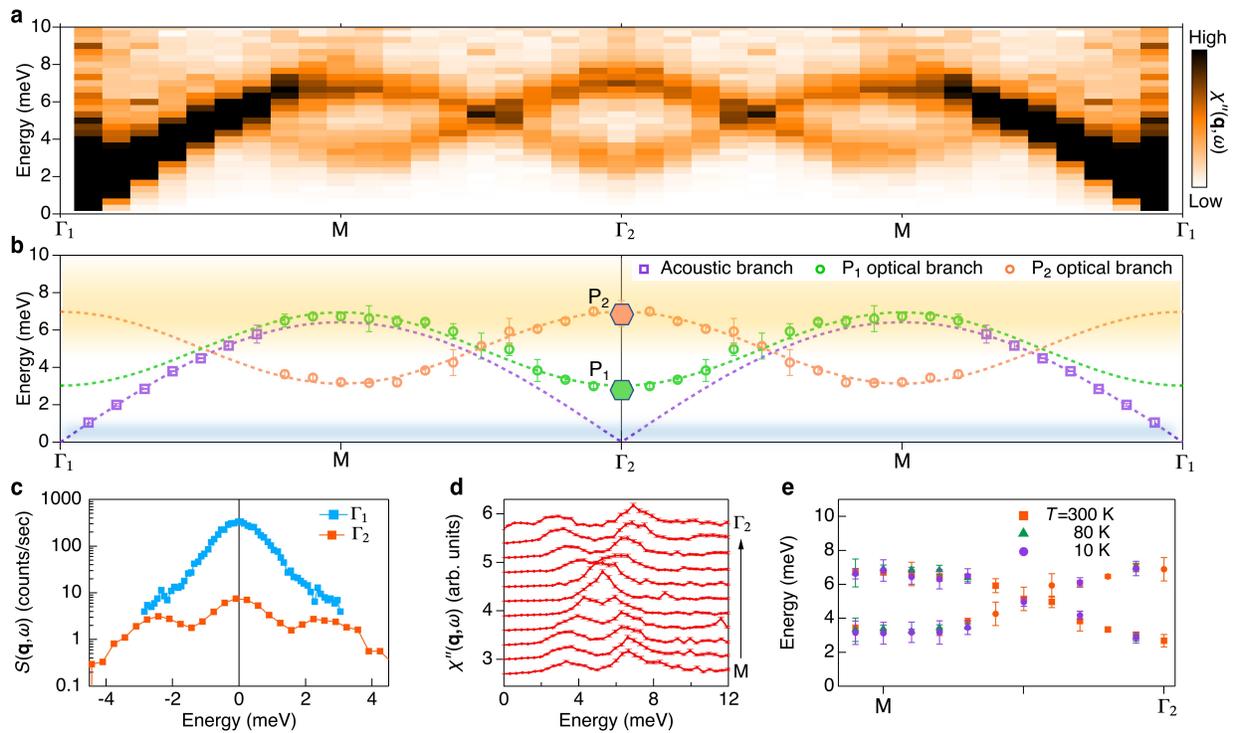

**Figure 2 Room temperature phonon excitations in α-RuCl₃.** **(a)** Low-energy phonon excitations determined by IXS along the $\Gamma_1$ (6, -3, 0)–M (6, -2.5, 0)–$\Gamma_2$ (6, -2, 0) direction. The plot shows the Bose-factor corrected IXS intensity. The extracted peak positions are presented in panel **(b)** revealing interlaced optical phonons intercepting with the transverse acoustic phonon branch. The optical phonon energies at $\Gamma_2$ are denoted by the green ($P_1$) and orange ($P_2$) hexagons, which are consistent with the phonon modes previously found by THz-spectroscopy[38]. The yellow

and blue shaded areas correspond to the two characteristic Kitaev energy scales displayed in Fig. 1(d). **(c)** IXS spectra at $\Gamma_1$ and $\Gamma_2$. The intensity is shown on a logarithmic scale. Note that due to the large intensity difference at $\Gamma_1$ and $\Gamma_2$, the acoustic phonon intensity is expected to be extremely weak near $\Gamma_2$. **(d)** Constant momentum transfer cuts around the phonon-crossing. The two optical branches cross each other without imposing a hybridization gap. **(e)** The extracted phonon peak positions from M to $\Gamma_2$ at different temperature reveals a temperature-independent massless Dirac-cone. The error bars in panel **b** and **e** denote the $2\sigma$ returned from the fittings (see Supplementary Note 2). The error bars in panel **d** represent one standard deviation assuming Poisson counting statistics.

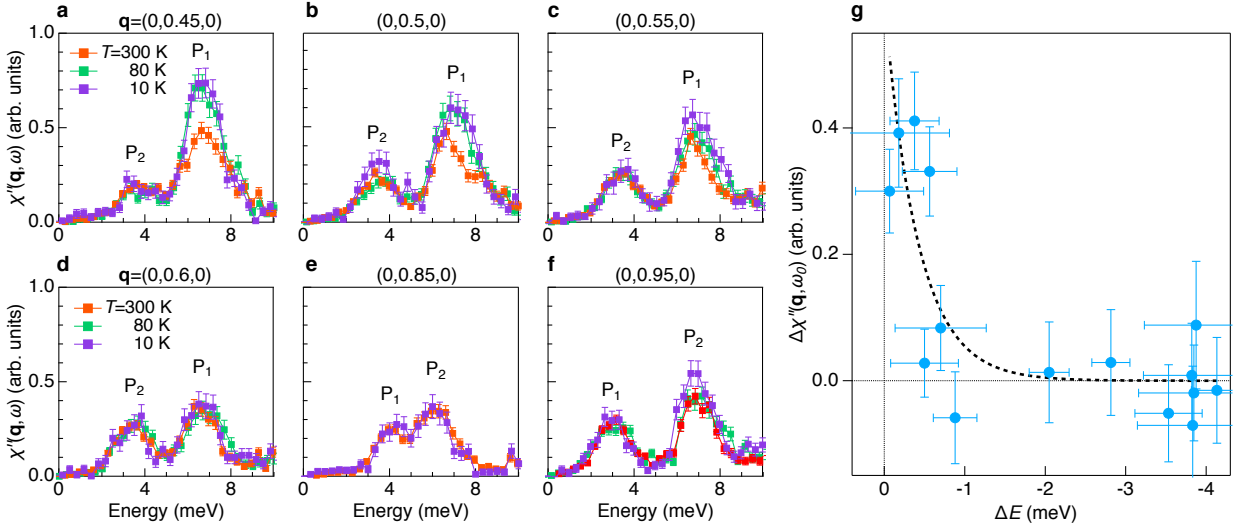

**Figure 3 Itinerant MF-phonon coupling near $\omega \sim J_K$. (a-f)** Spectra of the interlaced optical phonons at different reduced momentum transfer **q**. Here, we define **q**=(0,0,0) and (0,1,0) as $\Gamma_1$ and $\Gamma_2$ respectively, where the M point is at **q**=(0,0.5,0). The labels $P_1$ and $P_2$ denote the two optical phonon branches. Note the relative peak position of $P_1$ and $P_2$ switches at **q**=(0,0.75,0). The temperature dependent $\chi''(\mathbf{q}, \omega)$ shows a spectral weight enhancement at $\omega \sim J_K$ at low temperature. In panel **b**, we notice a shoulder on $P_2$ that may come from the acoustic mode. (**g**) The difference in the integrated phonon spectral weight, $\Delta\chi''(\mathbf{q}, \omega_0)$, between 10 and 300 K as function of $\Delta E = \omega_0 - \omega_{max}$. Here $\omega_0$ is the phonon peak position, $\omega_{max}$=7 meV is the band top of the interlaced optical phonons. The drastic increase of $\Delta\chi''(\mathbf{q}, \omega_0)$ is fitted to an exponential function (dashed line). The vertical error bars in all panels represent one standard deviation based on Poisson counting statistics. The horizontal error bars in panel **g** denote the $2\sigma$ returned from the fitting algorithm that extract the spectral peak positions.

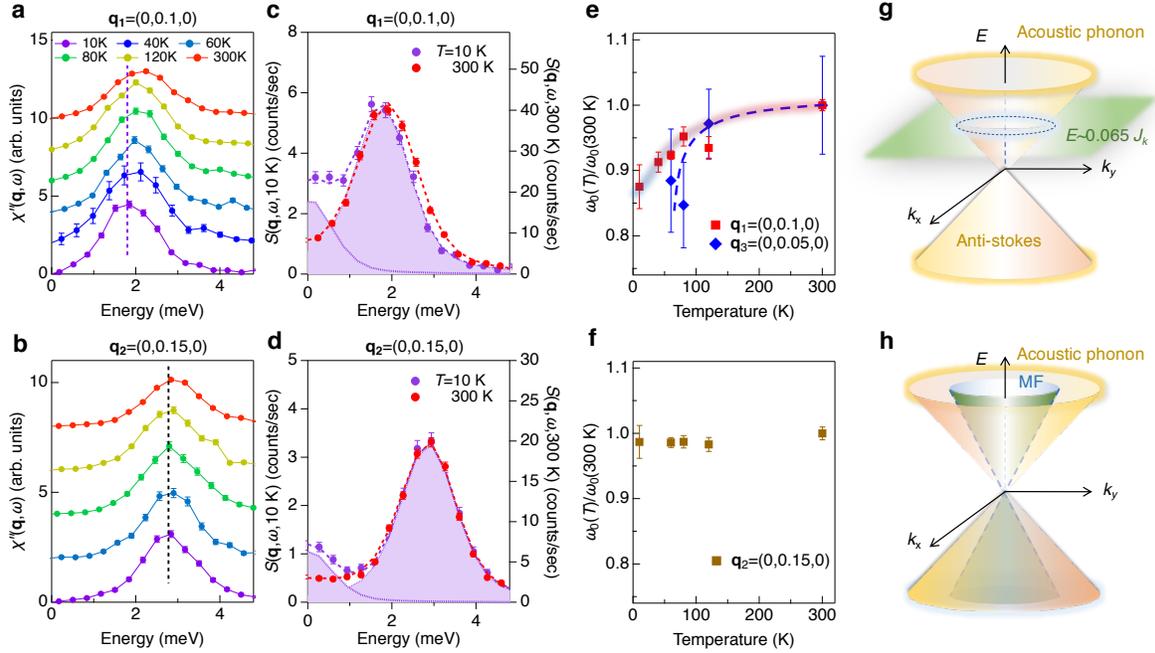

**Figure 4 Giant acoustic phonon softening.** (**a**) and (**b**) show the temperature-dependent $\chi''(\mathbf{q}, \omega)$ at $\mathbf{q_1}$=(0,0.1,0) in reciprocal lattice units (r.l.u.), $\omega \sim 2$ meV and $\mathbf{q_2}$=(0,0.15,0) r.l.u., $\omega \sim 3$ meV, respectively. (**c**) direct comparison of the IXS raw data, $S(\mathbf{q_1}, \omega)$, at $T$ =10 and 300 K. (**d**) shows

the same plot as (c) but at $q_2$. There is an apparent phonon softening at $q_1$, while at $q_2$, the effect is negligible. **(e-f)** The relative peak shift at $q_1$ and $q_2$. The ~13% phonon softening at $q_1$ (red squares in panel **e**) corresponds to a ~0.3 meV phonon peak shift. This value is as large as some well-known electron-phonon coupled systems[10]. The blue diamonds in panel **e** represent the relative peak shifts at $q_3$=(0,0.05,0) that show even larger softening-effect (~15% at 60 K), whereas $q_2$ display negligible change (panel **f**). This acoustic phonon anomaly, together with the spectral enhancement discussed in Fig. 3, present a full picture of the FPC in α-RuCl$_3$. Panel **g** and **h** schematically show two phonon coupling mechanisms. **(g)** The flatband of the $Z_2$ flux mode intercepts the acoustic phonon near $\omega$~0.065$J_K$. **(h)** The nearly identical linear dispersion of the itinerant MF and the acoustic phonon at $q \rightarrow 0$ causes a phonon renormalization at low temperature. The error bars in panels **a-d** represent one standard deviation assuming Poisson counting statistics. The error bars in panels **e-f** denotes the $2\sigma$ returned from the fitting.

**Acknowledgements:** We thank T. Berlijn, H. Ding, J. K. Keum, G. Kotliar, S. Nagler, N. Perkins, and A. Tennant for stimulating discussions. This research at Oak Ridge National Laboratory (ORNL) was sponsored by the U.S. Department of Energy, Office of Science, Basic Energy Sciences, Materials Sciences and Engineering Division (IXS data analysis, material synthesis and data interpretation) and by the Laboratory Directed Research and Development Program of ORNL, managed by UT-Battelle, LLC, for the U.S. Department of Energy (IXS experiment). Part of IXS data interpretation work at Brookhaven National Laboratory was supported by the U.S. DOE, Office of Science, Office of Basic Energy Sciences, Materials Sciences, and Engineering Division under Contract No. DE-SC0012704. This research used resources of the Advanced Photon Source, a U.S. Department of Energy (DOE) Office of Science User Facility, operated for the DOE Office of Science by Argonne National Laboratory under Contract No. DE-AC02-06CH11357. Extraordinary facility operations were supported, in part, by the DOE Office of Science through the National Virtual Biotechnology Laboratory, a consortium of DOE national laboratories focused on the response to COVID-19, with funding provided by the Coronavirus CARES Act. T.T.Z. and S.M. acknowledge the supports from Tokodai Institute for Element Strategy (TIES) funded by MEXT Elements Strategy Initiative to Form Core Research Center. T.T.Z. also acknowledge the support by Japan Society for the Promotion of Science (JSPS), KAKENHI Grant No. 21K13865. S.M. also acknowledges support by JSPS KAKENHI Grant No. JP18H03678. G.B.H. and S.O. were supported by the U.S. Department of Energy, Office of Science, National Quantum Information Science Research Centers, Quantum Science Center. D.G.M. acknowledges support from the Gordon and Betty Moore Foundation's EPiQS Initiative, Grant GBMF9069.